\documentstyle[aps,prl,epsf,twocolumn,floats]{revtex}

\begin{document}
\draft

\twocolumn[\hsize\textwidth\columnwidth\hsize\csname @twocolumnfalse\endcsname

\title{Frustration-induced non-Fermi-liquid behavior in a three-impurity
Kondo model}
\author{Bruce C. Paul and Kevin Ingersent\cite{byline}}
\address{Department of Physics, University of Florida,
P.\ O.\ Box 118440, Gainesville, FL 32611--8440}
\date{18 July 1996}
\maketitle

\begin{abstract}
A Kondo model for three spin-$\frac{1}{2}$ impurities placed at the vertices
of an equilateral triangle is studied using the numerical renormalization-group
method.
The impurity spins can form two frustrated doublets which, because of the
rotation symmetry, couple equally to the conduction band.
Over a broad range of antiferromagnetic RKKY interaction strengths, this
degeneracy drives the system to one of two novel, non-Fermi-liquid fixed
points.
Potential scattering is a relevant perturbation at one of these fixed points,
but is marginal throughout the second regime.
\end{abstract}


]
\narrowtext

Much of the rich behavior of heavy-fermion compounds \cite{Grewe} is thought
to stem from competition between ordering of local moments via the
Ruderman-Kittel-Kasuya-Yosida (RKKY) interaction, and quenching of those
moments due to the Kondo effect.
Theoretical understanding of the subject is far from complete.
It is widely believed that the essential physics is contained in the periodic
Anderson Hamiltonian, but this model has been solved only at the
mean-field level \cite{lattices}.
Insight has also been gained from exact solutions for one or two magnetic
impurities in simple metals \cite{Schlottman}, and for lattices in the
limit of infinite spatial dimensionality \cite{Georges}.

Recently, attention has focused on rare-earth and actinide systems which
exhibit a specific heat coefficient $C/T$ that diverges weakly with decreasing
temperature $T$ --- rather than remaining constant, as in a (heavy) Fermi
liquid --- and a resistivity that is linear --- instead of quadratic --- in $T$
\cite{vonLohneysen}.
Explanations proposed for these non-Fermi-liquid (NFL) properties include:
(1) the single-impurity Kondo effect with a disorder-induced distribution of
energy scales \cite{Dobrosavljevic};
(2) isolated magnetic impurities coupled to two or more
conduction bands \cite{multichan};
and (3) proximity to a $T=0$ magnetic-ordering transition \cite{QCP}.

A fourth possibility, overlapping with but not identical to (3) above,
is that lattice effects, such as periodicity or geometric
frustration, may induce NFL behavior where none exists in the one-impurity
limit.
Few-impurity systems may provide clues as to the plausibility of this scenario.
The two-impurity Kondo problem yields NFL physics ---
but only at particle-hole symmetry and with
fine-tuning of the RKKY coupling \cite{Jones,Affleck:92}.

In this paper, a more robust form of NFL behavior is identified in a model for
a single conduction band interacting with three identical spin-$\frac{1}{2}$
impurities arranged in an equilateral triangle.
Magnetic frustration prevents even strong antiferromagnetic RKKY interactions
from locking the impurities into a total spin singlet; instead, four
degenerate spin-$\frac{1}{2}$ configurations remain.
Using the numerical renormalization-group method, we find that over a wide
range of RKKY couplings, Kondo screening of these impurity states
is governed by two NFL fixed points.
Particle-hole asymmetry destabilizes one of the regimes, but is a
marginal perturbation in the other.
Our results suggest that spatial symmetry might give rise to NFL behavior
in more realistic models without the need for parameter tuning.
(Similar considerations have led to the study of electron-assisted
tunneling between three atomic sites \cite{Moustakis} as a possible
explanation for the anomalous conductance of certain point
contacts \cite{Ralph}.)

We start with a Hamiltonian, $H=H_{\text{\it band}}+H_{\text{\it int}}$,
describing a non-interacting conduction band coupled locally to three
spin-$\frac{1}{2}$ impurities, ${\bf S}_i$ ($i=0,1,2$), which lie at the
vertices ${\bf r}_i$ of an equilateral triangle.
The band is assumed to be isotropic in momentum space,
and to extend in energy over a range $\pm D$ about the Fermi level
(taken to be energy $\epsilon = 0$).
The interaction term is
\begin{equation}
   H_{\text{\it int}} = \sum_{i=0}^2 \left [ V n({\bf r}_i)
	       - J {\bf s}({\bf r}_i) \cdot {\bf S}_i \right] ,
							\label{H_start}
\end{equation}
where $n({\bf r}_i)$ [${\bf s}({\bf r}_i)$] is the electron number [spin] at
impurity site $i$.
The exchange coupling $J$ is assumed negative, whereas the potential
scattering $V$ can be of either sign.

Symmetry permits the impurities to couple to just six orthogonal, spatially
localized conduction states, annihilated by operators
$c_{h\sigma}$, where $\sigma=\pm\frac{1}{2}$ labels the spin
$z$ component, and $h\in\{0,1,2\}$ specifies the ``helicity.''
Helicity is analogous to parity in a two-impurity problem, i.e.,
a helicity-$h$ wave function acquires a multiplicative factor $e^{i2\pi h/3}$
under a rotation of $2\pi/3$ about the center of symmetry.
The combined states of three spin-$\frac{1}{2}$ impurities decompose into a
quartet having total spin $\frac{3}{2}$ and helicity $h=0$, plus two doublets
of total spin $\frac{1}{2}$.
The doublets can be constructed so that one has $h=1$ and the other $h=2$, in
which case the Hamiltonian conserves total helicity (modulo 3), and is
invariant under interchange of helicity labels $1$ and $2$.

Applying these observations, Eq.~(\ref{H_start}) can be rewritten
\begin{eqnarray}
   H_{\text{\it int}}
     &=& V_{0} n_{0} + V_{1} (n_{1} + n_{2})
        -J_{00}\,{\bf s}_{00}\!\cdot\!{\bf \tilde{S}}_0
							\nonumber \\
     & &-J_{11}({\bf s}_{11}+{\bf s}_{22})\!\cdot\!{\bf\tilde{S}}_0
        -J_{12}({\bf s}_{12}\!\cdot\!{\bf\tilde{S}}_1
              + {\bf s}_{21}\!\cdot\!{\bf\tilde{S}}_2)
							\nonumber \\
     & &-J_{01}[({\bf s}_{01} + {\bf s}_{20})\!\cdot\!{\bf\tilde{S}}_1
               +({\bf s}_{10} + {\bf s}_{02})\!\cdot\!{\bf\tilde{S}}_2 ] ,
							\label{H_int}
\end{eqnarray}
where
$n_{h}=\sum_{\sigma}c^{\dagger}_{h\sigma}c^{\rule{0ex}{1.35ex}}_{h\sigma}$
is the number of helicity-$h$ electrons at the impurity sites,
${\bf s}_{hh'}=\sum_{\sigma\sigma'}c^{\dagger}_{h\sigma} \frac{1}{2}
\bbox{\sigma}_{\sigma\sigma'} c^{\rule{0ex}{1.35ex}}_{h'\sigma'}$
gives the spin of an electron scattering from helicity $h'$ to $h$,
and ${\bf \tilde{S}}_{\Delta h} = \sum_{j=0}^2 e^{i2\pi j\Delta h} {\bf S}_{j}$ 
is the spin of the impurities as their helicity is raised by $\Delta h$.
The couplings
\begin{equation}
V_h = V F_h^2$, \quad $J_{hh'}= J F_h F_{h'}		\label{couplings}
\end{equation}
depend on the band dispersion $k(\epsilon)$, the density of states
$\rho(\epsilon)$, and the impurity separation $R$, through form factors
$
F_h^2 = \frac{1}{3} \int d \epsilon \, \rho(\epsilon)
	\left[ 1 + (3\delta_{h,0}-1)\sin (kR) / (kR) \right] .
$

This problem can be solved using a generalization of the non-perturbative
renormalization-group method developed for the one-impurity Kondo model
\cite{Wilson}.
The continuum of electronic energies is replaced by a discrete set,
$\pm D \Lambda^{-n}$, $n=0,1,\ldots$, where $\Lambda > 1$.
The Hamiltonian is then mapped, via the Lanczos method, onto a tight-binding
model for a one-dimensional chain with the impurities coupled to one end.
The discretization introduces a separation of energy scales which allows
the Hamiltonian to be solved iteratively, starting with a single site.
At each subsequent step, one site is added to the chain, while the effective
temperature $T$ falls by a factor of $\Lambda^{1/2}$.
This continues until the energy spectrum becomes scale-invariant, i.e.,
a stable fixed point is reached.

The discretized Hamiltonian contains an infinite set of parameters which
depend on the band shape and the impurity separation $R$.
We argue, along lines advanced for the two-impurity Kondo problem
\cite{Affleck:95}, that these band parameters can be replaced by standard
values calculated for $\rho(\epsilon)=$ constant and $R=\infty$.
All possible low-temperature regimes of the original model can be explored
provided that one treats the couplings entering Eq.~(\ref{H_int}) as
independent parameters, freed from the constraints implied by
Eqs.~(\ref{couplings}).
In particular, $J_{11}$ and $J_{12}$, although equal as originally defined,
should be regarded as separate couplings because no symmetry of the
problem prevents them from renormalizing independently.

We have solved the discretized version of this problem for different initial
values of the couplings $V_0$, $V_1$, $J_{00}$, $J_{11}$, $J_{12}$, and
$J_{01}$.
Even taking advantage of conserved quantities to break the Hamiltonian matrix
into block-diagonal form and thereby to reduce CPU time, we have generally
found it practical to retain no more than 1200 states after each iteration.
This corresponds to keeping all energies $E\lesssim 5T$, which allows
identification of fixed points, but is insufficient for direct computation
of physical properties as thermodynamic traces.
The good quantum numbers are total helicity, $H$;
total spin and its $z$ component, $S$ and $S_z$;
and total ``isospin'' \cite{Affleck:92} or ``axial charge'' \cite{Jones}
and its $z$ component, $I$ and $I_z$ \cite{charge}.
Total isospin is only conserved at particle-hole symmetry, so most runs were
performed with $V_0=V_1=0$ in order to benefit from the additional quantum
number.

The three-impurity Kondo problem has a rich phase diagram, as might be
expected given the large number of independent parameters.
In this paper, we focus on the stable low-temperature regimes of the
particle-hole-symmetric model.
Figure~\ref{fig:phase} shows a schematic phase diagram for this case,
projected onto the plane spanned by two dimensionless combinations of
the four exchange couplings entering Eq.~(\ref{H_int}):
The first, $K=(J_{00}^2 + 2J_{11}^2 - J_{12}^2 - 2J_{01}^2)/
(J_{00}^2 + 2J_{11}^2 + 2J_{12}^2 + 4J_{01}^2)$,
measures the strength of the RKKY interaction.
As $K$ ranges from $-\frac{1}{2}$ through $0$ to $+1$, 
the static impurity correlation function
$\langle\frac{1}{3}\sum_{i<j} {\bf S}_i\cdot{\bf S}_{j}\rangle$
varies smoothly from $-1/4$ through $0$ to $+1/4$.
The second combination,
$\alpha\equiv(J_{01}\!-\!J_{12})/(J_{01}\!+\!J_{12})$,
measures the difference between the two couplings that dominate the
region of antiferromagnetic RKKY interactions, $K<0$.

There are three distinct regions in Fig.~\ref{fig:phase}:

\begin{figure}
\centerline{
\vbox{\epsfxsize=65mm \epsfbox{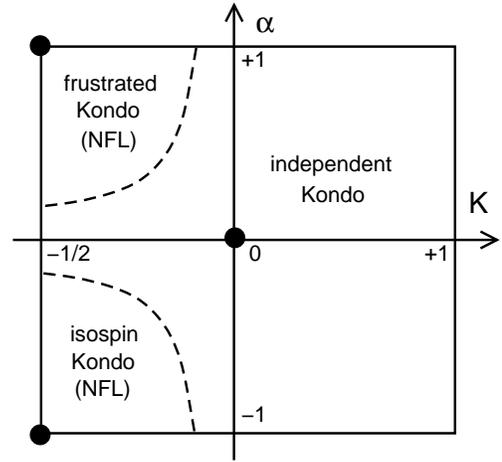}}
}
\vspace{2ex}
\caption{
Schematic phase diagram for the three-impurity Kondo model at particle-hole
symmetry.
Here $K$ is a measure of the RKKY interaction, and $\alpha$ measures
the relative strengths of the couplings $J_{01}$ and $J_{12}$
in Eq.~(\protect\ref{H_int}).
Circles represent stable fixed points, and dashed lines mark the borders
between different regimes of low-temperature behavior.
}
\label{fig:phase}
\end{figure}

{\em Independent-Kondo regime:\/}
For ferromagnetic and weakly antiferromagnetic RKKY couplings, the system
renormalizes to a stable fixed point at which interactions between the
impurities are essentially irrelevant.
The ground state is a spin singlet, and all electrons at the Fermi level
undergo a phase shift $\delta_h(\epsilon = 0) = \pi/2$.
Thus, the low-energy excitations are those for three impurities at infinite
separation, each independently undergoing a Kondo effect.
Particle-hole asymmetry is exactly marginal in this regime, just as it
is in the one-impurity Kondo model.

{\em Frustrated-Kondo regime:\/}
In a region of medium-to-strong antiferromagnetic RKKY interactions where
$J_{01}$ is the dominant coupling, the system flows to a novel,
stable NFL fixed point at which electrons of all three helicities screen
the frustrated spin-$\frac{1}{2}$ impurity configurations.

Table~\ref{table:frustrated} gives the finite-size spectrum at this fixed
point, computed for $\Lambda=3$, a value that represents a near-optimal
compromise between discretization errors (which grow with $\Lambda$) and
numerical rounding errors (which become larger as $\Lambda$ approaches unity)
\cite{Wilson}.
The energies have been multiplied by a factor of $0.625$, chosen to map the
smallest splitting in the spectrum of the free conduction band (without any
impurity) onto its value in the continuum limit, $\Lambda\rightarrow 1$.
The scaled energies $E$ are expressed in units of $\pi v_F/L$, where $L$
is the number of sites in the chain, and $v_F$ is the Fermi velocity.

\begin{table}
\renewcommand{\arraystretch}{1.1}
\[
\begin{array}[t]{c@{\hspace{0.15in}}l@{\hspace{0.05in}}l}
\multicolumn{3}{c}{L \text{ odd}} \\
E	  & \multicolumn{2}{c}{(S,I,H)} \\ \hline
0    	  & (0,0,0) 	& \times 2 \\
0.100	  & (\frac{1}{2},\frac{1}{2},*) \\
0.200	  & (0,1,*)	& (1,0,*) \\
0.299	  & (\frac{1}{2},\frac{1}{2},0) 	& \times 2 \\
0.499	  & (0,0,*) 	& \times 2 \\
0.599	  & (\frac{1}{2},\frac{1}{2},*) \\
0.601	  & (1,1,*) \\
0.604	  & (\frac{1}{2},\frac{1}{2},0) 	& \times 2 \\
0.699	  & (0,1,0)	& (1,0,0) \\
0.700	  & (0,1,0)	& (1,0,0) \\
0.701	  & (\frac{1}{2},\frac{3}{2},*)	& (\frac{3}{2},\frac{1}{2},*) \\
0.704	  & (0,1,*)	& (1,0,*) \\
0.804	  & \ldots
\end{array}
\qquad
\begin{array}[t]{c@{\hspace{0.15in}}l@{\hspace{0.05in}}l}
\multicolumn{3}{c}{L \text{ even}} \\
E	  & \multicolumn{2}{c}{(S,I,H)} \\ \hline
0	  & (0,\frac{1}{2},*)	& (\frac{1}{2},0,*) \\
0.400	  & (\frac{1}{2},1,*)	& (1,\frac{1}{2},*) \\ 
0.500	  & (0,\frac{1}{2},0)	& \times 2 \\
\text{''} & (\frac{1}{2},0,0)	& \times 2 \\
0.501	  & (0,\frac{1}{2},*)	& (\frac{1}{2},0,*) \\
0.603	  & (\frac{1}{2},1,0)	& \times 2 \\
\text{''} & (1,\frac{1}{2},0)	& \times 2 \\
0.799	  & (0,\frac{3}{2},0)	& (\frac{3}{2},0,0) \\
0.800	  & (0,\frac{3}{2},0)	& (\frac{3}{2},0,0) \\
0.904	  & (\frac{1}{2},1,0)	& (1,\frac{1}{2},0) \\
0.905	  & (\frac{1}{2},1,0)	& (1,\frac{1}{2},0) \\
0.913	  & (\frac{1}{2},1,*)	& (1,\frac{1}{2},*) \\
1.00	  & \ldots
\end{array}
\]
\caption{
Low-lying states from the finite-size spectrum at
the frustrated-Kondo fixed point.
Energies $E$ are given for odd- and even-length chains.
States are labeled by their spin $S$, isospin $I$, and helicity $H$;
an asterisk stands for one state each of helicity 1 and 2.
See the text for further details.
}
\label{table:frustrated}
\end{table}

The regularity of the levels in Table~\ref{table:frustrated} is reminiscent of
a Fermi liquid, but several details --- particularly the absence of states in
the odd-$L$ spectrum at $E\approx 0.4$, and the ordering of the even-$L$
states for $E<0.7$ --- prove impossible to explain within this framework.

The rate at which the eigenenergies approach their fixed-point
values also points to an NFL description:
By examining the $L$-dependence of the spectrum, and using the
relation $T\sim D\Lambda^{-L/2}$ \cite{Wilson}, we deduce that the leading
irrelevant operator at this fixed point varies like $T^\lambda$, with
$\lambda = 0.21 \pm 0.03$ (c.f. $\lambda=1$ for a Fermi liquid).
This exponent should be reflected in anomalous temperature dependences of
physical properties.

An NFL interpretation of the frustrated-Kondo regime is supported by evidence
that the fixed point occurs at finite, non-zero exchange couplings.
Empirically, one can eliminate all spin-$\frac{3}{2}$ impurity
configurations, and set $J_{00} = J_{11} = J_{12} = 0$ in Eq.~(\ref{H_int}),
to arrive at a one-parameter model which, for all $-D< J_{01}<+D$ (at least),
reproduces the fixed-point spectrum in Table~\ref{table:frustrated}.
It can be shown using poor-man's scaling \cite{Anderson} that the $J_{00}$
and $J_{11}$ terms in Eq.~(\ref{H_int}) are regenerated under
renormalization; however, $J_{12}$ remains identically zero.
In principle, one should be able to locate the fixed point by adjusting
$J_{00}$, $J_{11}$, and $J_{01}$ until the energy spectrum does not change
at all as the tight-binding chain is extended.
In practice, the band discretization introduces irrelevant operators which
renormalize the spectrum at small values of $L$.
We therefore associate the fixed point with the couplings which yield the
smallest deviation from the asymptotic spectrum at some standard chain
length, usually $L=20$.
This procedure gives $J_{00}^* = J_{01}^* = -0.8 D$ and
$J_{11}^* = +0.4 D$ (all to within $\pm 0.1D$), and $J_{12}^* \equiv 0$.

Throughout the frustrated-Kondo regime, potential scattering is found to
act as a marginal perturbation about the particle-hole symmetric limit, 
shifting the energy of each state by an amount proportional to its charge
quantum number, $I_z$.
Thus, for moderate values of $V_0$ and $V_1$ [see Eq.~(\ref{H_int})], this
regime retains its NFL character.
Very strong potential scattering suppresses the Kondo interaction, however,
and drives the system to a weak coupling Fermi liquid.
The crossover between these extremes is hard to analyze given the
limited number of states we can retain in our calculations, but this issue
merits future attention.

The evidence presented above demonstrates that the frustrated-Kondo
regime has an NFL fixed point which is stable with respect to the RKKY
interaction and is marginal under particle-hole asymmetry.
(However, breaking the real-space symmetry of the impurity locations will
lead to the recovery of Fermi-liquid behavior below some crossover
temperature.)
This fixed point does not seem to map onto any fixed point known from
previous studies of impurity models.
One likely candidate, the overscreened limit of the 3-channel Kondo
model for a spin-$\frac{1}{2}$ impurity, is ruled out because it predicts a
ground state spin $S=1$ for $L$ odd, and $S=\frac{1}{2}$ for $L$ even.

{\em Isospin-Kondo regime:\/}
In a second region of the parameter space --- characterized by
antiferromagnetic RKKY interactions, but dominated by $J_{12}$ rather
than $J_{01}$ --- the system flows to a stable NFL fixed point which
is related, by interchange of spin and isospin degrees of freedom,
to the one-impurity, two-channel Kondo model.

\begin{table}
\renewcommand{\arraystretch}{1.1}
\[
\begin{array}[t]{c@{\hspace{0.15in}}l@{\hspace{0.05in}}l}
\multicolumn{3}{c}{L \text{ odd}} \\
E	  & \multicolumn{2}{c}{(S,I,H)} \\ \hline
0    	  & (0,0,0)	& (0,1,0) \\
\text{''} & (\frac{1}{2},\frac{1}{2},0) \\
0.125	  & (0,0,*)	& (1,0,*) \\
\text{''} & (\frac{1}{2},\frac{1}{2},*) \\
0.500	  & (0,0,*)	& (0,1,*) \\
\text{''} & (\frac{1}{2},\frac{1}{2},0)	& (\frac{1}{2},\frac{1}{2},*) \\
\text{''} & (1,0,0)	& (1,1,0) \\
\text{''} & (\frac{3}{2},\frac{1}{2},0) \\
0.629	  & (0,1,*)	& (1,1,*) \\
\text{''} & (\frac{1}{2},\frac{1}{2},*)	& (\frac{1}{2},\frac{3}{2},*) \\
1.001	  & \ldots
\end{array}
\qquad
\begin{array}[t]{c@{\hspace{0.15in}}l@{\hspace{0.05in}}l}
\multicolumn{3}{c}{L \text{ even}} \\
E	  & \multicolumn{2}{c}{(S,I,H)} \\ \hline
0	  & (0,\frac{1}{2},0) \\
0.124	  & (\frac{1}{2},0,*) \\
0.498	  & (0,\frac{1}{2},*)	& (1,\frac{1}{2},0) \\
0.501	  & (\frac{1}{2},0,0)	& (\frac{1}{2},1,0) \\
0.625	  & (0,\frac{1}{2},*)	& (1,\frac{1}{2},*) \\
0.628	  & (\frac{1}{2},1,*) \\
0.997	  & (0,\frac{3}{2},0) \\
1.000	  & (\frac{1}{2},0,0)	& (\frac{1}{2},1,0) \\
\text{''} & (\frac{1}{2},0,*)	& (\frac{1}{2},1,*) \\
\text{''} & (\frac{3}{2},0,0)	& (\frac{3}{2},1,0) \\
1.003	  & \ldots
\end{array}
\]
\caption{
Low-lying states at the isospin-Kondo fixed point.
See Table~\protect\ref{table:frustrated} for an explanation of the entries.
}
\label{table:isospin}
\end{table}

Table~\ref{table:isospin} gives the finite-size spectrum at this fixed point,
once again computed for $\Lambda=3$ and scaled by a factor of $0.625$.
The energies and quantum numbers of all the low-lying states can be
reproduced by convoluting the free-fermion spectrum for decoupled helicity-$0$
electrons with a spectrum derived from the NFL fixed point of the one-impurity,
two-channel model \cite{Affleck:92a}:
For each state at the two-channel fixed point, labeled by spin $J$,
flavor $J_f$ and charge $Q$ \cite{Affleck:92a,flavor}, one constructs a
three-impurity state with the same energy, having quantum numbers
$S=J_f$; $I=S$; and $H=0$, $1$ or $2$,
depending on whether $Q=0$, $Q>0$ or $Q<0$.

Insight into these observations can be gained from another one-parameter
model.
Spin-$\frac{3}{2}$ impurity configurations are eliminated, just as for
the frustrated-Kondo fixed point, but here $J_{12}$ is the only non-zero
coupling.
For all $-D< J_{12}<+D$, this model flows to the fixed-point listed in
Table~\ref{table:isospin}.
Since $J_{00}=J_{01}=0$, the helicity-$0$ electrons must be decoupled at
the fixed point.
The interaction between the remaining degrees of freedom can be written
\begin{equation}
H_{\text{int}} = \frac{1}{4}\!
   \sum_{\mu,\nu=1}^3\sum_{h,h'=1}^2\sum_{\sigma,\sigma'}
   J_{\mu\nu} c^\dagger_{h\sigma} \sigma^\mu_{\sigma\sigma'} \sigma^\nu_{hh'}
   c^{\rule{0ex}{1.35ex}}_{h'\sigma'} S_\mu H_\nu,
							\label{double-tensor}
\end{equation}
where $\bf S$ and $\bf H$ are commuting SU(2) operators acting on the impurity
spin and helicity, respectively.
An isotropic ($J_{\mu\nu}=J$) version of Eq.~(\ref{double-tensor}) has been
considered in combination with pure spin and flavor interactions in a
generalized two-channel model \cite{Pang}, but we are aware of no study of
the case $J_{\mu\nu}=J(1-\delta_{\nu,3})$ which arises here.

Unlike the other two fixed points, the isospin-Kondo fixed point is
unstable with respect to particle-hole asymmetry.
Specifically, $V_1$ entering Eq.~(\ref{H_int}) plays the role of the
magnetic field in the standard two-channel Kondo model, driving the system to
a Fermi-liquid state.
Since there is no reason to expect $V_1$ to be zero in any real system,
this instability implies that the isospin-Kondo regime is unlikely to be
observed in practice.

In the preceding discussion, we have not specified the precise positions of the
boundaries between the three regimes shown in Fig.~\ref{fig:phase}.
These depend in a fairly complicated fashion on the two combinations
of exchange couplings orthogonal to $K$ and $\alpha$.
We note, though, that the frustrated-Kondo and isospin-Kondo regimes expand
as the Kondo coupling $J$ decreases in magnitude, i.e., as the RKKY coupling
grows to dominate the single-impurity Kondo effect.
We suspect that in the limit $J\rightarrow 0$, each NFL regime occupies an
entire quadrant of Fig.~\ref{fig:phase}.

We have presented numerical renormalization-group results for a model of three
magnetic impurities interacting with a non-magnetic metal.
For ferromagnetic and weakly antiferromagnetic RKKY interactions, the
impurities undergo essentially independent Kondo effects.
For stronger antiferromagnetic RKKY couplings, the three-fold
spatial symmetry of the model leads to two novel low-temperature regimes
of non-Fermi-liquid behavior, one of which is marginally stable against
particle-hole asymmetry.
Further work is required to identify the physical properties of the stable
non-Fermi-liquid regime, and to determine whether or not it can be reached
starting from any realistic band structure.

We thank B.\ Jones, A.\ Ludwig and A.\ Schiller for useful discussions.
This work was supported in part by NSF Grant No.\ DMR--9316587.


\begin{thebibliography}{18}
\bibitem{flavor}
    The generators of electron flavor symmetry are obtained from standard
    spin generators by interchanging spin and channel indices:
    e.g., $J_{f,z}$ measures the difference between the number of
    electrons in channels $1$ and $2$.
\end{thebibliography}
\end{document}